\documentclass[a4paper,11pt]{article}
\pdfoutput=1
\usepackage{jcappub}
\usepackage{amssymb,amsmath,mathrsfs,enumerate}
\usepackage{graphicx,rotate,multicol}
\usepackage{float}

\usepackage{subfig}

\usepackage{slashed}
\usepackage{mathtools}
\usepackage{multirow}

\allowdisplaybreaks

\title{\boldmath Primordial Black Holes from Slow Phase Transitions with Delayed Reheating: A Peak-Theory Approach}

\author[1,2]{Indra Kumar Banerjee}
\affiliation[1]{The Institute of Mathematical Sciences, Taramani, 600113 Chennai, India}
\affiliation[2]{Homi Bhabha National Institute, Training School Complex, Anushakti Nagar, Mumbai 400094,
India}

\emailAdd{indrakumarb@imsc.res.in}

\abstract{We study the possibility of significant PBH production from a slow first-order phase transition with delayed reheating. Since delayed reheating results in an early matter-dominated phase between percolation and reheating, we developed a peak-theoretic approach to PBH formation during this phase based on the non-Gaussian distribution of overdensity arising from the transition. To obtain the collapse probability, we performed large-scale Monte Carlo simulations and employed the hoop-conjecture criterion. We include tidal-torque terms to investigate the initial spin of the PBHs and find that the average spin parameter is $\mathcal{O}(10^{-3})$. Furthermore, we obtain an emergent overdensity threshold for collapse that depends on the phase transition properties and reheating efficiency. We find that the resulting PBH abundance is extremely sensitive to the reheating efficiency, with order-unity changes in efficiency leading to variations of many orders of magnitude in the collapse fraction. We identify regions of parameter space where the resulting PBHs can account for the entirety of the dark matter abundance. Finally, we also constrain the phase transition and reheating properties from current data on (non-)observations of PBHs.
}

\begin{document}
\maketitle
\flushbottom


\section{Introduction}
\label{sec:intro}
Primordial black holes (PBHs) were first proposed as compact objects that originated in the early universe from the collapse of overdense regions~\cite{Zeldovich:1967lct,Carr:1974nx}. Since PBHs can partially or completely account for all the dark matter (DM) in the universe~\cite{Carr:2009jm,Carr:2016drx,Carr:2017jsz}, a plethora of studies related to PBHs can be found in the literature (see Ref.~\cite{Carr:2020gox} for review). Although inflationary perturbations have traditionally been considered the source of the overdense regions required for PBH formation~\cite{Ivanov:1994pa}, various other mechanisms, such as cosmological first-order phase transitions (FOPTs)~\cite{Kodama:1982sf, Hsu:1990fg, Liu:2021svg, Hashino:2021qoq, Jung:2021mku, Kawana:2022olo, Lewicki:2023ioy, Gouttenoire:2023naa, Baldes:2023rqv, Gouttenoire:2023bqy, Salvio:2023ynn, Gouttenoire:2023pxh, Jinno:2023vnr, Banerjee:2023qya, Flores:2024lng, Lewicki:2024ghw, Lewicki:2024sfw, Ai:2024cka, Conaci:2024tlc, Kanemura:2024pae, Banerjee:2024cwv, Hashino:2025fse, Murai:2025hse, Banerjee:2024fam, Kawana:2021tde, Dent:2025lwe}, domain walls~\cite{Deng:2016vzb, Deng:2017uwc, Deng:2020mds, Gouttenoire:2023gbn}, and cosmic strings~\cite{Hawking:1987bn, Polnarev:1988dh}, have also been considered. Recently, FOPT-sourced PBHs have gained significant traction from both model-dependent and independent perspectives, as FOPTs can generate correlated PBHs and gravitational waves (GWs)~\cite{Caprini:2015zlo, Caprini:2019egz, Ellis:2019oqb, Banerjee:2023brn, Banerjee:2023vst}. Among these, the creation of PBHs from density perturbations due to the delayed decay of the false vacuum in a slow FOPT is particularly interesting, as various beyond the standard model (BSM) scenarios can accommodate such FOPTs. These studies showed that sufficiently slow and strong FOPTs can generate enough PBHs to account for the entire DM of the universe.  

However, a few studies questioned this conclusion due to inconsistencies in gauge choice. Specifically, it was shown in Refs.~\cite{Franciolini:2025ztf, Wang:2026zvz} that the high overdensity tail of the density perturbation distribution is significantly larger in the previously used spatially flat gauge in comparison to the comoving gauge; however, the previous studies used the threshold value of the overdensity, considered as the criterion of the PBH creation, in the comoving gauge, which led to an overestimation of the PBH abundance by many orders of magnitude. The authors of Ref.~\cite{Franciolini:2025ztf} further showed that once this gauge issue is fixed, even the slowest FOPTs are rendered insufficient for generating significant PBHs. On this note, recently, Ref.~\cite{Ai:2026zrs} argued that slow FOPTs can be revived as an efficient PBH-generating event if one considers non-instantaneous reheating at the end of the FOPT. The primary novel aspect of the mechanism outlined in Ref.~\cite{Ai:2026zrs} is that, between percolation and reheating, condensates of the FOPT driving scalar dominate the universe, giving rise to an early matter-dominated (EMD) universe. In contrast to radiation domination, in EMD, due to the lack of pressure of the background fluid, perturbations can grow linearly with the scale factor. Therefore, even small overdensities can form PBHs. Furthermore, Ref.~\cite{Ai:2026zrs} claims the possibility of near-extremal initial spin of the resulting PBHs. 

In this article, we explore this mechanism in detail, employing a peak-theory approach to estimate the abundance and spin of PBHs originating in EMD. In particular, we incorporate the full deformation tensor, including the off-diagonal tidal terms and Hessian eigenvalues, and perform Monte Carlo simulations to estimate the likelihood of collapse of an overdense region in EMD. Some of the main improvements of our approach are that we identify an emergent density threshold specific to the underlying FOPT parameters, obtain an extended mass function, and determine the initial spin distribution. We find that the hoop conjecture preferentially selects near-spherical regions to collapse, leading to very small initial spins of the PBHs. Finally, we show the sensitivity of the present PBH properties to the reheating efficiency.      

The article is organized as follows: in Sec.~\ref{sec:outline} we provide a brief outline of the density perturbations arising from slow FOPT, followed by Sec.~\ref{sec:pbh_emd} where we discuss the creation of PBHs in EMD, specifically how PBHs originate from a slow FOPT with delayed reheating, and discuss the numerical methods employed. In Sec.~\ref{sec:prop_pbh}, we discuss the initial PBH properties, and in Sec.~\ref{sec:res}, we present the results. Finally, in Sec.~\ref{sec:conc} we summarize and conclude.
\section{Outline of the existing idea}
\label{sec:outline}
In this section we briefly discuss the basics of a slow and supercooled FOPT, the generation of density perturbations and delayed reheating at the end of the FOPT.
\subsection{Recap on slow and supercooled FOPTs}
\label{subsec:fopt}
Slow and supercooled FOPTs are parametrized by two important quantities, i.e., (i) $\beta/H$, which is a measure of the inverse duration of the FOPT in terms of the Hubble time at the FOPT epoch, and (ii) $\alpha$, which encodes the ratio of the vacuum energy and radiation energy densities at the FOPT epoch. The effective FOPT epoch is defined by the percolation time $t_*$ (or conversely the percolation temperature $T_*$), i.e., the time when only $71\%$ of the volume of the universe remains in false vacuum. 

Since a FOPT progresses through the nucleation and subsequent expansion of true vacuum bubbles, mathematically, the dynamics of the FOPT are primarily controlled by the nucleation rate, which can be expressed as,
\begin{align}
\Gamma = \Gamma_0\exp(\beta(t - t_n)),
\end{align}
where $t_n$ is the nucleation time, i.e., the time at which the probability of a true vacuum bubble to exist per Hubble volume per Hubble time is unity. After the nucleation time, it becomes dynamically favorable for true vacuum bubbles to nucleate, and multiple true vacuum bubbles form until they collide and percolate at $t_*$. In a slow and supercooled FOPT, due to the domination of vacuum and the relatively longer duration of the FOPT, the universe briefly goes through vacuum domination until the expanding bubbles of the FOPT driving scalar \textit{sweeps} most of the regions with false vacuum at $t_*$. At this point, if the decay rate of the scalar $\Gamma_{\phi} > H_*$, then the universe reheats and goes into radiation domination almost immediately, whereas in the case of $\Gamma_{\phi} < H_*$, the reheating is inefficient and the universe expands in an effective matter domination (caused by the scalar condensate) till $H$ decreases. The reheating occurs at $t_{\rm reh} (> t_{*})$, which is defined by $H(T_{\rm reh}) = \Gamma_{\phi}$. Therefore, the comoving Hubble radius $r_{\mathcal{H}} = \mathcal{H}^{-1} = (aH)^{-1}$ reduces during vacuum domination and then expands as the vacuum domination ends, where $a$ and $H$ are the scale factor and the Hubble parameter. Since the vacuum domination ceases at $t_*$, the largest comoving wavenumber (corresponding to the smallest length scale) that exits the horizon is $k_{\rm max} = a(t_*)H(t_*) = a_*H_*$. This scale is the first to re-enter the horizon after percolation, and then as time progresses, other scales with $k = a(t_k)H(t_k)<k_{\rm max}$ re-enter the horizon at time $t_k$. Depending on the efficiency of the reheating, the immediate epoch of the universe could either be RD or EMD, and in either case, the $t_k$ can be expressed as,

\begin{align}
	t_k &= \begin{cases} 
        \dfrac{1}{2H_k} = \dfrac{1}{2H_*}\left(\dfrac{k}{k_{\rm max}}\right)^{-2} & \text{for RD, i.e, instantaneous reheating,}~~ \Gamma_{\phi}/H_* \geq 1 \\
        \dfrac{2}{3H_k} = \dfrac{2}{3H_*}\left(\dfrac{k}{k_{\rm max}}\right)^{-3}  & \text{for EMD, i.e, delayed reheating,}~~ \Gamma_{\phi}/H_* < 1  \\
        \end{cases}
\end{align}

Nucleation of a true vacuum bubble is a probabilistic process; therefore, in different Hubble patches, it might be slightly earlier or later. If nucleation is delayed, the patch stays in vacuum domination for longer than the surrounding regions. This naturally generates density perturbations. In general, this density perturbation can be expressed as,

\begin{align}
\delta = \dfrac{\delta\rho}{\rho_{\rm avg}} = \dfrac{\delta\rho_V + \delta \rho_{r,s}}{\rho_{V, \rm avg} + \rho_{r,s, \rm avg}},
\end{align}
where the suffix $V,~r,~s$ stands for vacuum, radiation, and scalar; the quantities in the denominator with the suffix `avg' denote the average background values, whereas the quantities in the numerator provide $k$-mode dependent perturbations. In the case of instantaneous (delayed) reheating, we consider the radiation (scalar) energy density. Initially, i.e., at $t_n$, $\rho_V = \Delta V$ and $\rho_{r,s} = 0$, where $\Delta V$ is the difference in energy density between false and true vacuum.

As the bubbles nucleate and expand, the average fraction of the universe in false vacuum reduces; this can be expressed as,
\begin{align}
 F_{\rm avg}(t) = \exp\left[-\frac{4\pi}{3}\int_{-\infty}^{t} dt^{\prime}\Gamma(t^{\prime})\left(\int^{t}_{t^{\prime}}dt^{\prime\prime}\frac{a(t^{\prime})}{a(t^{\prime\prime})}\right)^{3}\right],
\end{align}
where the bubble wall velocity $v_w \sim 1$ and the average vacuum energy density can be expressed as $\rho_{V, \rm avg}(t) = F_{\rm avg}(t)\Delta V$. However, due to the probabilistic nature of the bubble nucleation, there could be mode-dependent fluctuations in the energy density on top of this average value, i.e., $\delta\rho_{V} = (F_k - F_{\rm avg})\Delta V$. Through cosmological perturbation theory, one can then study the evolution of the density perturbation $\delta$ for various $k$-modes up to horizon re-entry in various gauges, such as the flat gauge, comoving gauge, Newtonian gauge, etc. For further details of the gauge dependence of this quantity, see Refs.~\cite{Franciolini:2025ztf, Ai:2026zrs} (and references therein). In this study, we focus on this quantity in the comoving gauge and discuss the implications.

\subsection{Density perturbations}
\label{subsec:den_pert}
\begin{figure}[H]
\centering
\includegraphics[scale=0.55]{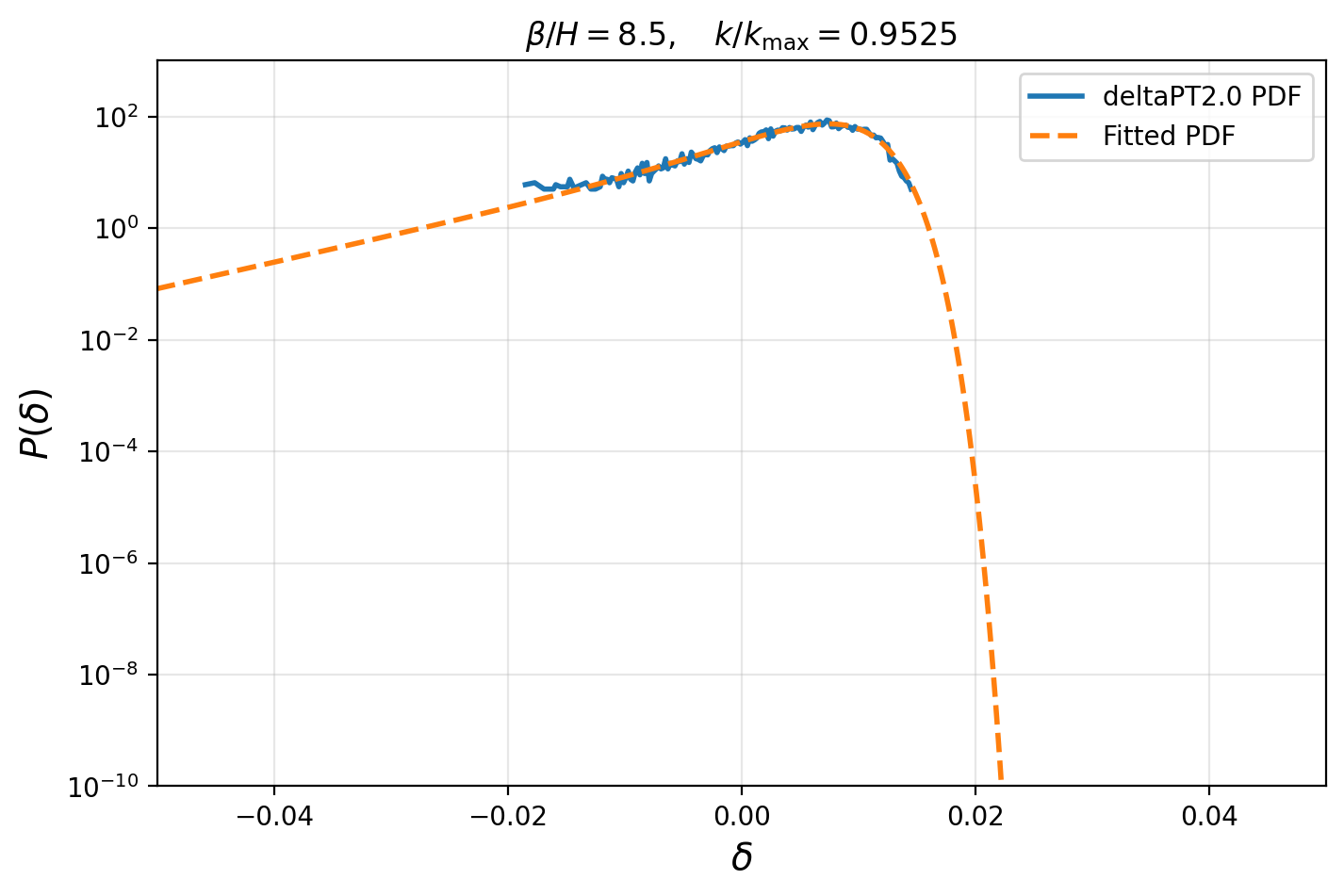}
\caption{The blue curve denotes the PDF of $\delta$ from the MCMC of \texttt{deltaPT2.0} and the orange curve denotes the fitting function for the same for $\beta/H = 8.5$ and $k/k_{\rm max} = 0.9525$.}
\label{fig:pdelta}
\end{figure}

As mentioned previously, the density perturbation in the comoving gauge is of specific interest to us, as it encodes information regarding the possibility of PBH creation from FOPTs. Using \texttt{deltaPT2.0}~\cite{Lewicki:2024ghw, Franciolini:2025ztf} we generate the $k$-mode dependent PDF of $\delta$ ($P(\delta)$) for $\beta/H \in (4,20)$, $j_c = 50$, and $N_{\rm sim} = 10^5$. We found that, for all cases, $k/k_{\rm max} \lesssim 0.8$ cannot generate enough overdensities to result in significant PBH production. Hence, in this study, we constrain ourselves to $k/k_{\rm max} \in (0.8, 1)$. Furthermore, since the PBH production is an extremely rare event and the usual realistic probabilities of PBH production are many orders of magnitude less than realistic values of $1/N_{\rm sim}$, one has to use a numerical fit of the PDF $P(\delta)$ to proceed with further analysis. We use the following fitting function~\cite{Franciolini:2025ztf},
\begin{align}
	P(\delta) = P_0 \exp\left[\frac{\epsilon}{2}(\delta - \mu) - \frac{2}{\epsilon^2\sigma^2}\left(1 - e^{\frac{\epsilon}{2}(\delta - \mu)}\right)\right],
\end{align}   
where the parameters $\mu,~\sigma, ~\epsilon(>0)$ denote the mean, the width and the skewness, i.e., the non-Gaussianity. As an illustrative example, we show the PDF obtained from \texttt{deltaPT2.0} and the fitted function for $\beta/H = 8.5$ and $k/k_{\rm max} = 0.9525$ in Fig.~\ref{fig:pdelta}.
The figure above shows that the PDF is highly non-Gaussian. Furthermore, the PDF's skewness leads to minuscule probabilities of high overdense regions. Therefore, in RD, where the threshold for the overdensity in the comoving gauge for PBH production $\delta_{\rm th, RD} \in [0.4, 0.67]$~\cite{Escriva:2021aeh, Musco:2018rwt, Escriva:2019phb}, formation of PBHs in significant abundance becomes impossible. Hence, for models where $\Gamma_{\phi}\geq H_*$, production of PBH for any cosmological significance is effectively ruled out. However, for models with $\Gamma_{\phi} < H_*$, there exists an EMD phase before reheating~\cite{Ellis:2019oqb}; during this EMD phase, depending on the duration of the phase, even smaller overdensities can eventually collapse to form PBHs. In the subsequent parts of this article, we discuss the dynamics of overdensities in the EMD phase and the possibility of PBH formation, depending on the phase duration.

\subsection{Delayed reheating}
\label{subsec:del_reh}
As mentioned before, in supercooled FOPTs, the vacuum energy dominates before percolation. In case of instantaneous reheating, one finds that immediately after percolation the universe reheats to a temperature $T_*$, where the vacuum energy density is related to this temperature as,
\begin{align}
\Delta V = \frac{\pi^2}{30}g_*T_*^4,
\end{align}
where $g_*$ is the number of relativistic degrees of freedom at $T_*$. However, this relation is not valid for $\Gamma_\phi/H_* < 1$, in which case, the reheating temperature can be approximately estimated by~\cite{Hambye:2018qjv},
\begin{align}
T_{\rm reh} \approx 0.64 \left(\frac{\Gamma_{\phi}}{H_*}\right)^{1/2}T_* .
\end{align} 
Furthermore, the relation between the reheating time and the percolation time can be expressed as,
\begin{align}
\frac{t_{\rm reh}}{t_*} = \frac{H_*}{\Gamma_{\phi}}.
\end{align}
Evidently, the quantity $\Gamma_{\phi}/H_*$ controls the duration of the EMD epoch between the percolation and the reheating. In the next section, we discuss how the collapse time differs depending on the shape of the overdensities in the EMD epoch.

\section{Creation of PBHs in EMD}
\label{sec:pbh_emd}
In the usual cases of PBH formation from FOPT, instantaneous reheating is assumed, and the PBH forms in RD. The initial PBH formation probability can be estimated simply from $\beta_{x_k, \rm RD} \sim \int_{\delta_{\rm th, RD}}^{\infty}P(\delta,x_k)$, where $x_k = k/k_{\rm max}$. In MD, this picture changes significantly, as we discuss below.

The density and any existing angular momentum of the overdense regions grow with time in EMD, and, as mentioned before, the collapse threshold of the overdensity is much lower than that in the case of RD. However, there are several other factors that enter the PBH creation prescription, such as the non-sphericity of these regions, the distribution of non-spherical regions, the hoop conjecture, etc., and we give a brief outline of those below.

\subsection{Evolution and collapse of density perturbations}
In this part of the article, we provide a brief overview of the formalism we use to obtain the dynamical evolution and eventual collapse of the density perturbation during the EMD phase of the universe~\cite{Ye:2025wif}.
\subsubsection*{Zel'dovich approximation}
To present the overall dynamics of the expanding background universe, i.e., to track comoving quantities such as radius, volume, etc., one uses Eulerian coordinates. However, the creation of PBHs requires tracking the evolution of rare fluid elements, and in that case, Lagrangian coordinates take precedence. In a Lagrangian coordinate system, a fluid element's initial position vector is labeled with $q_i$, and its Eulerian counterpart can be expressed as~\cite{Zeldovich:1969sb},
\begin{align}
x_i(\vec{q},t) \approx q_i - \tilde{D}_{ij}q_j,
\end{align}
where $\tilde{D}_{ij} \propto \partial^2\psi/\partial q_i\partial q_j$ is the gravitational tidal tensor with $\psi$ as the gravitational potential. This is the Zel'dovich approximation that relates the time-dependent Eulerian and the constant Lagrangian coordinates. It is to be noted that $\tilde{D}_{ij}\propto a(t)$, i.e., it grows during the EMD phase of the universe. 

In this article, we adopt the peak-theory approach to PBH formation, i.e., we assume that PBHs form from rare overdense regions. These regions have a peak overdensity $\delta_{\rm peak}$ enclosed within them, and the boundary of such a region can be expressed through the condition $\delta(q_{\rm boundary}) = 0$. It can be considered without any loss of generality that $\delta(\vec{q} = 0) = \delta_{\rm peak}$ and the overdensity field can be expanded around this as~\cite{Ye:2025wif},
\begin{align}
\delta = \delta_{\rm peak} - \frac{1}{2}\sigma_2\sum_{i = 1,2,3}\lambda_{i}q_i^2,
\end{align}
where $\lambda_{i}$ are the eigenvalues of the rescaled Hessian matrix $\delta_{ij}/\sigma_0$ with $\delta_{ij} = \partial^2\delta/\partial q_i\partial q_j$. The quantities $\sigma_i$ are the spectral moments of the power spectrum of the density perturbation. 

The peak-statistics formalism employed in this work requires knowledge of the spectral moments of the density perturbation field. However, since our framework is fundamentally based on the statistical distribution of the overdensity perturbations generated during the first-order phase transition, the underlying mode-dependent density power spectrum is not directly available. To overcome this issue, we construct an effective density perturbation spectrum, denoted by $\mathcal{P}_{{\rm eff},\delta}(k) = \int d\delta (\delta - \mu)^2 P(\delta|k)$, whose spectral moments, defined as $\sigma_i^2(k_0) = \int d\ln k k^{2i}\mathcal{P}_{{\rm eff},\delta}(k) W^2(k, k_0)$, reproduce the variance structure and local peak statistics required in the collapse formalism. Here $W^2(k,k_0) = \exp(-\ln^2(k/k_0)/\Delta^2)$ is the log-normal window function with $\Delta$ controls the width of the window function. We used $\Delta = 0.05$, which is an extremely narrow smoothing kernel, so, effectively, the spectral moments in this case are probing the local properties around the scale $k_0$. We emphasize that $\mathcal{P}_{{\rm eff},\delta}(k)$ should be interpreted as an effective statistical description of the density fluctuations rather than a uniquely reconstructed primordial power spectrum. 

With the foundations laid, we now move to the dynamics of gravitational collapse in EMD.

\subsubsection*{\textit{Pancake} Collapse}
Unlike in RD, in EMD, non-spherical overdensities retain their non-sphericity due to the absence of pressure from the background fluid. As a result, the overdensities can be ellipsoidal, and the different axes of this ellipsoid may collapse at different times. If non-sphericity is too high, the shortest axes might collapse much earlier, leading to a \textit{pancake}-like object, and this mechanism is termed the pancake collapse. The evolution of such an ellipsoidal region can be expressed as~\cite{Ye:2025wif},
\begin{align}
r_i(t) = a(t)x_i(t) = a(t)T_{ij}(t)q_j,
\end{align}
where the matrix $T_{ij} = \mathbf{1}_{ij} - \sigma_0(t)\tilde{D}_{ij}/\sigma_0$, where $\tilde{D}_{ij}/\sigma_0$ is the rescaled deformation matrix, $\mathbf{1}$ is the identity matrix, $\sigma_0(t) \propto a(t)$ and $\sigma_0(t_k) = \sigma_0$. For a general case, the principal axes co-ordinate of the rescaled Hessian matrix does not coincide with the principal axes of the rescaled deformation matrix and the latter can therefore be parametrized as~\cite{Ye:2025wif},
\begin{align}
\frac{\tilde{D}_{ij}}{\sigma_0} = 
\begin{bmatrix} 
\frac{1}{3}\tilde{D}_A + \tilde{D}_B + \frac{1}{3}\tilde{D}_C & -w_3 & -w_2 \\ 
-w_3 & \frac{1}{3}\tilde{D}_A - \frac{2}{3}\tilde{D}_C & -w_1 \\ 
-w_2 & -w_1 & \frac{1}{3}\tilde{D}_A - \tilde{D}_B + \frac{1}{3}\tilde{D}_C 
\end{bmatrix},
\end{align}
where $\tilde{D}_A = \delta/\sigma_0$ parametrizes the overdensity of a region, $\tilde{D}_{B,C}$ denotes the anisotropic collapse velocity along the principal axes, and $\omega_{1,2,3}$ denotes the tidal torque which gives a measure of the angular momentum in the overdense region. In accordance with the existing literature, we define $\nu = \delta/\sigma_0$ which, in Gaussian peak-theory, simultaneously characterizes the peak amplitude and its statistical rarity. 

Without the loss of generality, one may choose a point within the ellipsoidal region with Lagrangian coordinates $\overline{q}_{i}$ which diagonalizes the rescaled deformation matrix with the eigenvalues $d_{1,2,3}$.
One can then track the evolution of the three principal axes in the comoving Eulerian coordinates as,
\begin{align}
r_1(t) &= a(t)(1 - \sigma_0(t)d_1)\overline{q}_1,\\
r_2(t) &= a(t)(1 - \sigma_0(t)d_2)\overline{q}_2,\\
r_3(t) &= a(t)(1 - \sigma_0(t)d_3)\overline{q}_3.
\end{align}
For each axis, initially, when $\sigma_0$ is small, the fluid expands with the universe. If $d_i > 0$, then at $t_{i,\rm ta}$, where $\sigma_0(t_{i,ta}) = 1/(2d_i)$, the fluid along this direction stops expanding; this is known as the `turn around' point. Finally, at $t_{i,c}$ where $\sigma_0(t_{i,c}) = 1/d_i$, the fluid along that direction collapses. On the other hand, if $d_i < 0$, then throughout the EMD, that direction expands.
It can be understood from the above evolution equations that if $d_{\rm max} = \text{max}(d_1, d_2, d_3) = d_i$, then the $i$-direction will collapse first. In this formalism, we consider this time to be the universal collapse time of the overdense region; hence, we define $t_{c} = 1/d_{\rm max}$. At this point, since the collapse has occurred in one direction, one is left with an elliptical region.
At $t = t_{c}$ one can estimate the dimensionless geometric properties, such as the squared dimensionless semi-major and semi-minor axes of the elliptical region from the non-zero eigenvalues of the matrix $T(t_c)\Lambda^{-1}T(t_c)$ where $\Lambda = \text{diag}(\lambda_1, \lambda_2, \lambda_3)$. The non-zero eigenvalues can be denoted as $x^2_{l,s}$ where $x_{l} > x_{s}$, i.e., $x_{l}~(x_{s})$ is the dimensionless semi-major (minor) axes of the elliptical pancake at the time of collapse. Furthermore, the eccentricity of the ellipse can be expressed as $e = (1 - x_s^2/x_l^2)^{1/2}$.

However, mere collapse along one direction does not guarantee the formation of a BH; the hoop conjecture provides a threshold for a region to form a BH, and we discuss the conjecture as follows.

\subsubsection*{Hoop conjecture} 
Hoop conjecture states that \textit{black holes with horizons form when and only when a mass $M$ gets compacted into a region whose circumference in every direction is $\mathcal{C} \lesssim 4\pi M$} (in units where $G = c = 1$)~\cite{Misner:1973prb}. In order to visualize, it can be stated that for an imploding non-spherical region with mass $M$, collapse into a BH is only possible when an imaginary hoop of circumference $\mathcal{C}$ can be put around it and can be rotated $2\pi$ radians without any part of the imploding region coming out of the hoop. Therefore, $\mathcal{C}$ is the maximum circumference of the region in all directions if the region has to form a BH. Since at $t = t_c$, the maximum `length' of the region from the center of the region is $x_l$, the hoop conjecture criterion can be expressed as~\cite{Ye:2025wif},
\begin{align}
x_l \lesssim x_{\rm th} = \frac{\pi d_{\rm max}}{2E(e)(\lambda_1\lambda_2\lambda_3)^{1/6}}\sigma_0,
\label{eq:hoop}
\end{align}
where $x_{\rm th}$ is the dimensionless radius of the hypothetical hoop and $E(e)$ is the complete elliptical integral of the second kind. Only regions satisfying this criterion are considered to be collapsed to form a BH.

\subsubsection*{Spin of collapsing region}
An important aspect of non-spherical collapse is the angular momentum of the collapsing region. Angular momentum is generated due to the misalignment between the tidal tensor and the moment of inertia tensor. For a spherical region, the principal axes of these tensors coincide, leading to vanishing angular momentum. However, in the case of ellipsoidal regions, collapsing regions could have high angular momentum. Since we are interested in the properties of the resulting PBHs, we focus on the dimensionless spin parameter $a_* = S/M^2$ where $S$ and $M$ are the total angular momentum and mass of the collapsing region. One can express the squared spins in each of the three directions at $t = t_c$ as~\cite{Ye:2025wif},
\begin{align}
a_*^2(t_c) \approx \frac{2}{25}\frac{\lambda_1\lambda_2\lambda_3}{\delta d_{\rm max}^3}\left(\omega_1^2\left(\frac{1}{\lambda_2}-\frac{1}{\lambda_3}\right)^2 + \omega_2^2\left(\frac{1}{\lambda_1}-\frac{1}{\lambda_3}\right)^2 + \omega_3^2\left(\frac{1}{\lambda_2}-\frac{1}{\lambda_1}\right)^2\right). 
\label{eq:spin_para}
\end{align}  
From the above expression, it is evident that a collapsed pancake will have an extremely high spin parameter if $\delta \ll 1$. However, as discussed in the next part of the article, the hoop conjecture acts as a severe filter: only near-spherical regions with reasonably low peak parameter values satisfy the criterion, leading to low spin for the resulting PBHs. In the subsequent part of the article, we discuss the distributions of the quantities defining the overdense region properties such as $\nu,~\tilde{D}_{B},~\tilde{D}_{C},~\omega_{1,2,3},~\lambda_{1,2,3}$ and eventually the fraction of the overdense regions which satisfy the hoop criterion.

\subsubsection*{Joint distribution of peak variables}

In Gaussian peak-theory, one determines the joint probability distributions of the quantities under the assumption that all of them arise from the same underlying mechanism, i.e., the gravitational potential. Under this assumption, one can determine the number of peaks in parameter ranges from $\nu,~\tilde{D}_{B},~\tilde{D}_{C},~\omega_{1,2,3},~\lambda_{1,2,3}$ to $\nu + d\nu,~\tilde{D}_{B} + d\tilde{D}_{B},~\tilde{D}_{C}+d\tilde{D}_{C},~\omega_{1,2,3}+d\omega_{1,2,3},~\lambda_{1,2,3}+d\lambda_{1,2,3}$ per comoving unit volume as~\cite{Bardeen:1985tr},
\begin{align}
n_{\rm peak}^{\rm Gaussian}d\nu d\tilde{D}_B d\tilde{D}_C d^3\omega_{1,2,3} d^3\lambda_{1,2,3} &= A\left(\frac{\sigma_2}{\sigma_1}\right)^3 \exp\left(-\frac{Q_3}{2}\right)\lambda_{1}\lambda_2\lambda_3\nonumber\\
&\times(\lambda_2 - \lambda_3)(\lambda_1 - \lambda_3)(\lambda_1 - \lambda_2)d\nu d\tilde{D}_B d\tilde{D}_C d^3\omega_{1,2,3} d^3\lambda_{1,2,3},
\end{align}
where,
\begin{align}
Q_3 &= x^2 + \frac{(\nu - x\gamma)^2}{1-\gamma^2} + 15y^2 + \frac{(\tilde{D}_B - \nu\gamma)^2}{1-\gamma^2} + 5z^2 + \frac{5(\tilde{D}_C - z\gamma)^2}{1-\gamma^2} + 15\frac{\omega_1^2 + \omega_2^2 + \omega_3^2}{1-\gamma^2},\\
A &= \frac{3^{11/2}5^{5}}{2^{13/2}\pi^{11/2}(1-\gamma^2)^3},\\
x &= \lambda_1 + \lambda_2 + \lambda_3,\\
y &= \frac{\lambda_1 - \lambda_3}{2},\\
z &= \frac{\lambda_1 - 2\lambda_2 + \lambda_3}{2},\\
\gamma &= \sqrt{\frac{\sigma_1^2}{\sigma_2\sigma_0}},
\end{align}
and the added physical conditions are $\lambda_1 \geq \lambda_2 \geq \lambda_3$ and $\nu > 0$.

However, in our case, $\nu$ follows a non-Gaussian distribution. Although we have the distribution of $\delta$, i.e., $P(\delta)$ for fixed values of $\beta/H$ and $k/k_{\rm max}$, in each case, we find a fixed value of $\sigma_0$, and hence we construct $P(\nu)$ for each case. Since the non-Gaussian statistics of the remaining variables are not presently available, we assume that the conditional distributions of $\tilde{D}_{B,C},~\omega_{1,2,3}, \lambda_{1,2,3}$ retain their Gaussian peak-theory form. Under this assumption, the joint distribution becomes,
\begin{align}
n_{\rm peak}^{\rm NG}(\nu, \tilde{D}_{B,C}, \omega_{1,2,3}, \lambda_{1,2,3}) = P(\nu)n_{\rm peak}^{\rm Gaussian}(\tilde{D}_{B,C}, \omega_{1,2,3}, \lambda_{1,2,3}|\nu).
\end{align}
Equivalently, this may be viewed as a reweighting of the Gaussian peak distribution by the ratio $P(\nu)/P_{G}(\nu)$ ($P_G(\nu)$ is the Gaussian distribution of $\nu$) while retaining the Gaussian conditional distributions of the remaining peak variables.

\subsection{Numerical simulations and fitting functions}

To determine the probability of PBH formation from a given overdensity realization, we performed a large-scale Monte Carlo analysis over the space of local peak configurations. We scanned over our regime of interest, i.e., $\beta/H \in (4,20)$ and $k/k_{\rm max} \in (0.8,1)$ and found that the variables $\delta$ and $\sigma_0$ lie in range $(10^{-3}, 10^{-1})$ and $(10^{-5}, 10^{-3})$ respectively for significant values of $P(\nu)$. Furthermore, we also found that in this range for all the cases $\gamma \in (0.9975, 0.999)$. As a result of our subsequent analysis, we fix $\gamma = 0.998$. Upon this finding, we constructed a 2D grid of dimensions $50 \times 20$ in $(\delta, \sigma_0)$ within the ranges mentioned above. For each pair $(\delta,\sigma_0)$, corresponding respectively to the coarse-grained overdensity amplitude and the variance of the smoothed density field, we generated an ensemble of stochastic peak realizations characterized by the local Hessian eigenvalues $\lambda_i$, isotropic and anisotropic deformation variables $\tilde{D}_{A,B,C}$, and rotational degrees of freedom $\omega_i$. 

The Hessian eigenvalues determine the local curvature structure of the overdensity peak and therefore control the degree of spherical symmetry of the configuration. The isotropic deformation variable $\tilde{D}_A = \nu$ denotes the overdensity of the collapsing region, the anisotropic deformation variables $\tilde{D}_{B,C}$ characterize the quadrupolar distortion of the collapsing region, while the $\omega_i$ variables determine the local rotational support generated through mode coupling and anisotropic collapse. Collapse is therefore not solely controlled by the overdensity amplitude $\delta$, but also by the detailed geometric structure of the peak.

For each point in the $(\delta,\sigma_0)$ plane, we generated $10^7$ independent realizations of the stochastic variables. The sampled realizations were then tested against a collapse criterion derived from the hoop conjecture outlined in Eq.~\eqref{eq:hoop}. Configurations exhibiting excessive anisotropy, large quadrupolar deformation, or sufficiently strong rotational support fail to satisfy the hoop criterion and do not form PBHs.

The collapse fraction was then estimated as the fraction of realizations satisfying the hoop criterion,
\begin{equation}
F(\delta,\sigma_0)
=
\frac{N_{\rm coll}}{N_{\rm MC}},
\end{equation}
where $N_{\rm coll}$ denotes the number of successful collapsing realizations and $N_{\rm MC}$ is the total number of Monte Carlo samples. Repeating this procedure over the dense grid in the $(\delta,\sigma_0)$ plane yielded the full collapse kernel $F(\delta,\sigma_0)$.

In addition to determining the collapse probability, we also extracted the dimensionless spin parameter $a_*$ (Eq.~\eqref{eq:spin_para}) and the collapse time $t_{c}$ for each successful realization. Averaging over all collapsing configurations allowed us to construct the functions
\begin{equation}
a_*(\delta,\sigma_0),
\qquad
t_{c}(\delta,\sigma_0),
\end{equation}
which were subsequently used in the statistical evolution of the PBH properties.

Since PBH formation constitutes a rare-event process, naive Monte Carlo sampling becomes highly inefficient near the collapse threshold, where the allowed collapsing region occupies only a tiny fraction of the full configuration space. In these regions, direct sampling would lead to extremely poor statistics and large fluctuations in the estimated collapse fraction. To overcome this issue, we implemented an importance-sampling procedure that preferentially samples collapse-favorable configurations.

Instead of sampling directly from the original probability distribution $n_{\rm peak}^{\rm NG}(\nu, \tilde{D}_{B,C},$ $\lambda_{1,2,3},\omega_{1,2,3})$, we introduced a biased proposal distribution $q(\nu, \tilde{D}_{B,C},\lambda_{1,2,3},\omega_{1,2,3})$ that enhances the sampling of approximately spherical, low-spin, and weakly deformed configurations, which dominate the collapsing subset of the ensemble. Each realization was then assigned the standard reweighting factor
\begin{equation}
w
=
\frac{n_{\rm peak}^{\rm NG}}{q},
\end{equation}
thereby preserving the unbiased nature of the estimator. The collapse fraction was therefore computed using the weighted estimator
\begin{equation}
F(\delta,\sigma_0)
=
\frac{1}{N_{\rm MC}}
\sum_{i=1}^{N_{\rm MC}}
w_i\,\Theta_i,
\end{equation}
where $\Theta_i=1$ for realizations satisfying the collapse criterion and vanishes otherwise.

Importance sampling substantially reduced the variance of the Monte Carlo estimator and enabled us to accurately estimate the exponentially suppressed collapse probability in the near-threshold regime. This was particularly important because the final PBH abundance is exponentially sensitive to the detailed behavior of the collapse kernel.

The resulting Monte Carlo outputs consisted of large numerical datasets for the collapse fraction $F(\delta,\sigma_0)$, the average spin $a_*(\delta,\sigma_0)$, and the collapse time $t_{c}(\delta,\sigma_0)$. In order to enable efficient parameter scans over the cosmological and reheating parameter space, these numerical datasets were subsequently compressed into analytical surrogate fitting functions, which can be expressed as,
\begin{align}
F(\delta,\sigma_0) &\approx \frac{1}{1 + \left(\dfrac{0.55855\sigma_0^{0.49852}}{\delta}\right)^{9.661}},\\
a_{*}(\delta,\sigma_0) &\approx \frac{0.01856\sigma_0^{0.248}}{1+\left(\dfrac{\delta}{0.54473\sigma_0^{0.4807}}\right)^{2.2874}},\\
\frac{t_c}{t_k}(\delta, \sigma_0) &\approx 51.0187\sigma_0^{1.4969}\delta^{-3.0012} \approx 51.02\left(\frac{\sqrt{\sigma_0}}{\delta}\right)^3.
\end{align}
These fits formed the basis of the statistical kernel used throughout the remainder of the analysis.

Along with these, we also obtained the fitting functions for $d_{\rm max}$ and $\lambda_1$ for successful PBH formation realizations. These can be expressed as,
\begin{align}
d_{\rm max}(\delta,\sigma_0) &\approx 0.3404\sigma_0^{-0.998}\delta^{1.004 + 0.0009\log\sigma_0} \approx 0.34\frac{\delta}{\sigma_0},\\
\lambda_1(\delta,\sigma_0) &\approx 0.2912 \sigma_0^{-1.012}\delta^{0.9875} \approx 0.2912\frac{\delta}{\sigma_0}.
\end{align}
The function $d_{\rm max}$ acts as a consistency check as existing studies find $d_{\rm max} \sim 0.33(\delta/\sigma_0) + \mathcal{O}(\sigma_0)$ which is in excellent agreement with our findings~\cite{Ye:2025wif}. 
\section{Initial PBH properties}
\label{sec:prop_pbh}
From the observational aspects, the three PBH properties which we are most interested in are (i) PBH mass, (ii) PBH abundance, and (iii) initial spin of the PBHs. Therefore, the determination of these quantities is the main objective of this section.

\subsection*{Mass}

In this mechanism, we obtain an extended mass function of the resulting PBHs. The mass of these PBHs can be fixed by fixing the scale of the FOPT, i.e., $T_*$. Since $k_{\rm max} = a_*H_*$ and $(M/M_{\rm min}) = (k/k_{\rm max})^{-3}$, one can fix the mass range of the PBH population arising from this mechanism through fixing $M_{\rm min}$, which can be expressed as~\cite{Carr:2018nkm, Carr:2016hva},
\begin{align}
M_{\rm min} = \kappa M_{H_*},
\end{align}   
where $\kappa$ is the efficiency factor of the collapse. In this case we consider $\kappa \sim \mathcal{O}(1)$. On the other hand, $M_{H_*}$ is the horizon mass at percolation, which can be expressed as,
\begin{align}
M_{H_*} \approx 9.8\times 10^{32}\mathrm{~g}\left(\frac{g_*}{106.75}\right)^{-1/2}\left(\frac{T_*}{\rm GeV}\right)^{-2}.
\end{align}

\subsection*{Abundance}
As mentioned above, the suppression factor due to the hoop conjecture is encoded in the quantity $F(\delta|\sigma_0(x_k, \beta/H))$. Furthermore, the ratio of the collapse time and the percolation time can be expressed as
\begin{align}
\frac{t_c}{t_*}(\delta|\sigma_0(x_k, \beta/H)) = \frac{t_c}{t_k}(\delta|\sigma_0(x_k, \beta/H))(x_k)^{-2}.
\end{align} 
Hence, for a given $x_k$ and $\beta/H$, the initial collapse fraction can be expressed as,
\begin{align}
\beta_{\rm PBH} = \int_{0^{+}}^{\infty} d\delta P(\delta)F(\delta)\Theta\left(\frac{t_{\rm reh}}{t_*} - \frac{t_c}{t_k}(\delta)x_k^{-2}\right).
\end{align}
Unlike RD, in this case the collapse threshold of the overdensity $\delta_c$ emerges dynamically from the condition $t_{\rm reh} \geq t_c(\delta)$. For a fixed $\beta/H$ value, it depends on $x_k$ and $\Gamma_{\phi}/H_*$. The threshold can be expressed as,
\begin{align}
\delta_c \approx \left(\frac{51.02\sigma_0(x_k, \beta/H)^{3/2}(\Gamma_{\phi}/H_*)}{x_k^2}\right)^{1/3}
\label{eq:deltac}
\end{align}
Once the initial collapse fraction is obtained, one can estimate the present abundance of the PBHs as,
\begin{align}
f(M/M_{\rm min}) &= \frac{1}{\Omega_{\rm DM}}\beta_{\rm PBH}(M/M_{\rm min})\frac{T_{\rm reh}}{T_{\rm eq}},\\
& = \left(\frac{\beta_{\rm PBH}(M/M_{\rm min})}{3.25\times 10^{-8}}\right)\left(\frac{T_*}{\rm GeV}\right)\left(\frac{\Gamma_{\phi}/H_*}{10^{-4}}\right)^{1/2},
\end{align}
where $\Omega_{\rm DM} = 0.26$~\cite{Planck:2018vyg} is the relative density of the dark matter in the universe and $T_{\rm eq} = 0.8~ \rm eV$~\cite{dodelson2020modern} is the temperature of the universe at matter-radiation equality.
\subsection*{Spin}
As discussed in the previous section, the average spin of a PBH can be obtained from the peak parameter values of the collapsing overdense region. Therefore, for fixed values of $\beta/H$ and $x_k$, we obtain the spin parameter averaged over the peak variables. This is a statistical average of the spin parameter, and for fixed $\beta/H,~x_k$ it can be expressed as,
\begin{align}
\langle a_* \rangle = \frac{1}{\beta_{\rm PBH}}\int_{0^{+}}^{\infty} d\delta P(\delta)F(\delta)a_{*}(\delta)\Theta\left(\frac{t_{\rm reh}}{t_*} - \frac{t_c}{t_k}(\delta)x_k^{-2}\right).
\end{align}

\section{Results}
\label{sec:res}
In this section we discuss the dependence of the PBH properties on the FOPT variables, such as $\beta/H$, $\Gamma_{\phi}/H_*$, and $T_*$. 

\subsection{Role of $\beta/H$ and $\Gamma_{\phi}/H_{*}$}
\begin{figure}[H]
\centering
\includegraphics[scale=0.32]{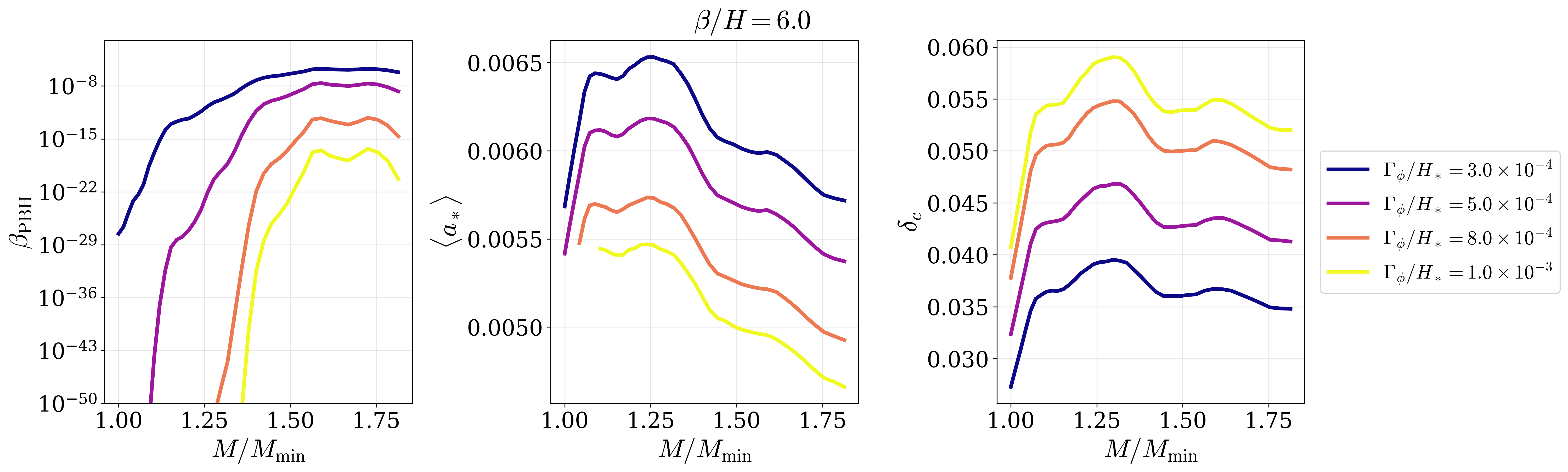}\\
\includegraphics[scale=0.32]{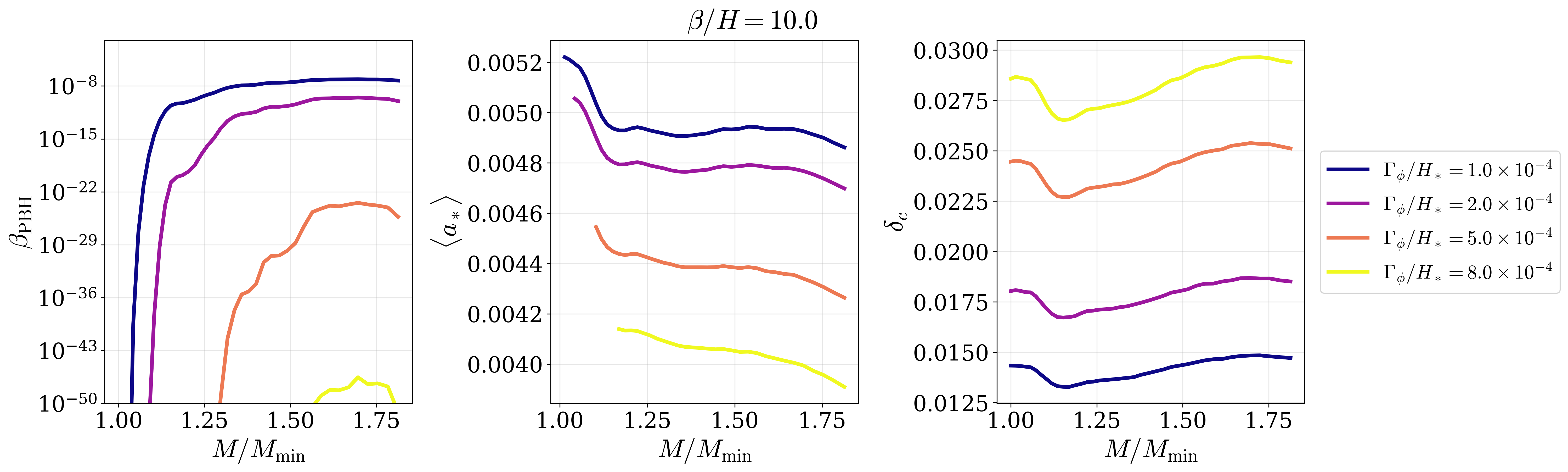}\\
\includegraphics[scale=0.32]{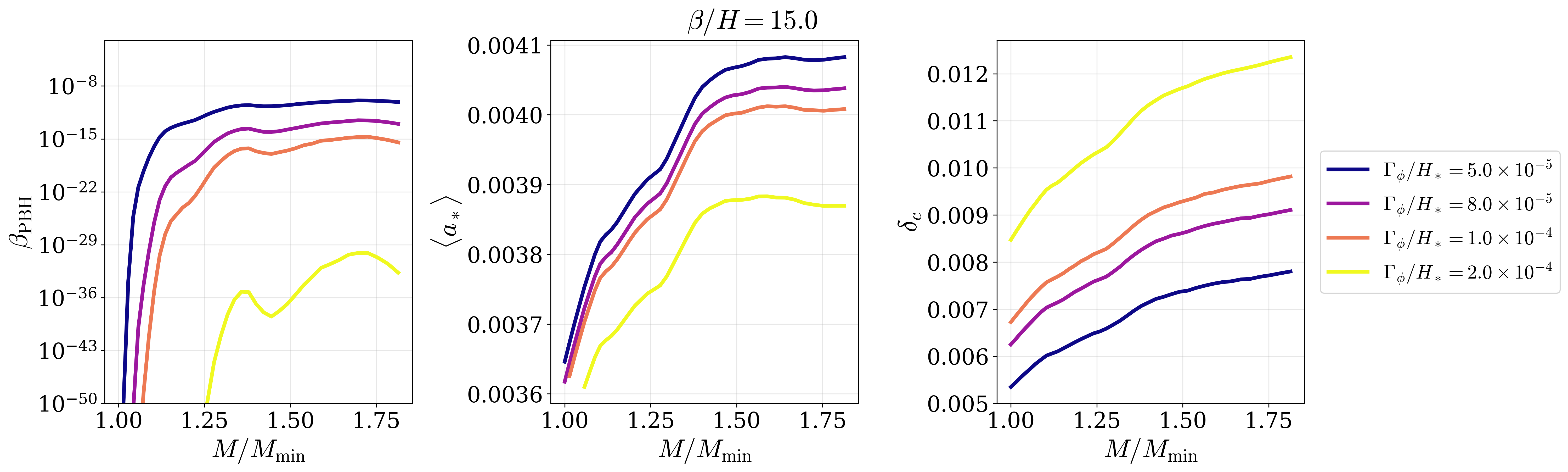}

\caption{Behavior of $\beta_{\rm PBH}$, $\langle a_* \rangle$ and $\delta_c$ in the relevant mass range corresponding to $x_k \in (0.8, 1.0)$ for different values of $\beta/H$ and $\Gamma_{\phi}/H_*$.}
\label{fig:betahgammapl}
\end{figure}

Since the duration of the FOPT and subsequent EMD is controlled by $\beta/H$ and $\Gamma_{\phi}/H_*$ respectively, the initial PBH-related quantities depend on these two variables as follows:
\begin{itemize}
\item \textbf{Duration of FOPT - $\beta/H$:} The duration of the FOPT is inversely proportional to the quantity $\beta/H$. Therefore, for fixed $\Gamma_{\phi}/H_*$ and larger $\beta/H$ fewer patches go through delayed nucleation, leading to a suppression in the fraction of the regions with large overdensity. The dependence of the PBH-related quantities are as follows:
\begin{enumerate}
\item \textbf{Initial collapse fraction - $\beta_{\rm PBH}$:} For smaller values of $\beta/H$ the overdensity PDF $P(\delta)$ has higher values for larger overdensities than the same in case of larger $\beta/H$. As a result, a larger number of regions which satisfy the hoop conjecture criterion can grow and collapse to form PBHs before $t = t_{\rm reh}$. Hence, $\beta_{\rm PBH}$ grows as $\beta/H$ reduces.
\item \textbf{Collapse threshold - $\delta_c$:} It can be seen from Eq.~\eqref{eq:deltac} that for fixed $\Gamma_{\phi}/H_*$, $\delta_c \propto \sigma_0^{1/2}$ and smaller values of $\beta/H$ increase $\sigma_0$. As a result, the collapse threshold increases as $\beta/H$ decreases. In peak-theory, the absolute value of overdensity alone is not sufficient to characterize a peak, one needs to consider $\nu$ which denotes the statistical significance or rarity of a region. In this case, for larger $\sigma_0$ values, regions with the same $\delta$ become less `exceptional' and therefore take longer time to collapse. Since $\delta_c$ is a threshold determined by the condition $t_c \leq t_{\rm reh}$, smaller $\beta/H$ leads to larger $\sigma_0$ which in turn increases the collapse threshold. However, the effect of threshold increment arising from smaller $\beta/H$ has insufficient effect on $\beta_{\rm PBH}$ as that is largely controlled by the dramatic increment of regions with larger overdensities.
\item \textbf{Average spin - $\langle a_* \rangle$:} From the Monte Carlo realizations encoded in the fitting function $F(\delta, \sigma_0)$ we find that for fixed $\sigma_0$, larger $\delta$ leads to rapid increase in the fraction of overdensity that satisfies the hoop conjecture criterion. In other words, extremely rare peaks ($\nu \gg 1$) tend to be more spherical. As a result, rarer peaks are less prone to acquire high angular momentum before collapse. In this work we see that the collapse threshold scales as $\delta_c \propto \sigma_0^{1/2}$ which leads to the scaling of the threshold rarity parameter $\nu_c = \delta_c/\sigma_0 \propto \sigma_0^{-1/2}$. Since slower FOPTs correspond to larger $\sigma_0$ values, larger $\beta/H$ leads to collapse of exceptionally rare regions, leading to smaller angular momentum, thus smaller average spin parameter. However, we find that in the relevant range of $\beta/H$ that we consider in this work, the strong filtering of hoop conjecture only allows spin $\mathcal{O}(10^{-3})$, which is consistent with Ref.~\cite{Ye:2025wif}.
\end{enumerate}

\item \textbf{Duration of EMD - $\Gamma_{\phi}/H_*$:} The duration of the EMD phase is inversely proportional to $\Gamma_{\phi}/H_*$. Therefore, for a fixed $\beta/H$ value, smaller values of $\Gamma_{\phi}/H_*$ facilitates the possibility of collapse of very small overdensities. The dependence of PBH-related quantities are as follows:
\begin{enumerate}
\item \textbf{Initial collapse fraction - $\beta_{\rm PBH}$:} Since prolonged EMD phase allows much smaller overdensities to collapse, smaller values of $\Gamma_{\phi}/H_*$ leads to larger collapse fraction.
\item \textbf{Collapse threshold - $\delta_c$:} For fixed $\beta/H$, hence fixed $\sigma_0$, the collapse threshold $\delta_c \propto (\Gamma_{\phi}/H_*)^{1/3}$. This is again due to the reason that longer EMD allows smaller overdensities to grow large enough for collapse. Hence, smaller $\Gamma_{\phi}/H_*$ corresponds to smaller collapse threshold.
\item \textbf{Average spin - $\langle a_* \rangle$:} Longer EMD phase allows the tidal torque to act for a longer duration making it possible for the overdense regions to acquire larger angular momentum, and therefore larger average spin. Furthermore, as mentioned above, a prolonged EMD phase decreases $\nu_c$ allowing for less rare (and less spherical) peaks to collapse and the increase in non-sphericity results in larger average spin.  Hence, smaller $\Gamma_{\phi}/H_*$ corresponds to larger $\langle a_* \rangle$.
\end{enumerate}

\end{itemize}
In order to illustrate this further, we show $\beta_{\rm PBH},~\langle a_* \rangle,~\delta_c$ for the relevant mass range for different $\beta/H$ and $\Gamma_{\phi}/H_*$ in Fig.~\ref{fig:betahgammapl}.
The above figure shows the behaviour of $\beta_{\rm PBH},~\langle a_* \rangle,~\delta_c$ with $\beta/H$, and $\Gamma_{\phi}/H_*$ as explained above. One of the major aspects of this is that although in this study we include the tidal terms in the calculation, the value of the average spin $\langle a_* \rangle \sim \mathcal{O}(10^{-3})$. This is due to the fact that regions with higher angular momenta are filtered by the hoop conjecture criterion, as mentioned above. Furthermore, in the same parameter region, the collapse fraction changes by many orders of magnitude, whereas spin only changes at the percent level. 

Finally, we point out that for the different $\beta/H$ values the shape of the curves is different with varying $M/M_{\rm min}$. This non-triviality is sourced by the fact that $P_{\mathrm{eff}, \delta}(k)$ depends non-trivially on $x_k$ for different values of $\beta/H$ as the different modes enter at different times and experience fluctuations differently.
\subsection{Role of $T_{*}$}
\begin{figure}[H]
\centering
\includegraphics[scale=0.22]{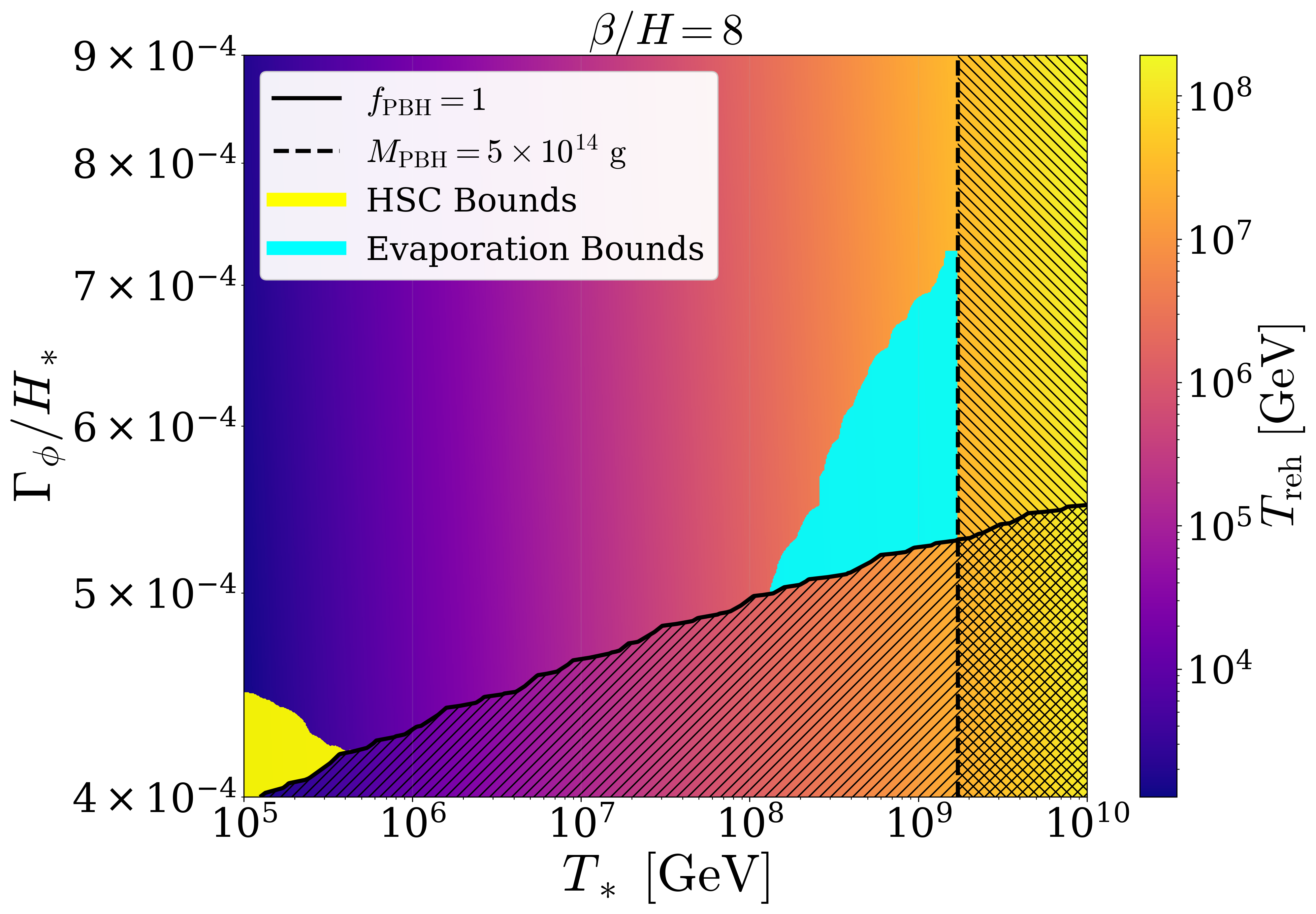}~~
\includegraphics[scale=0.22]{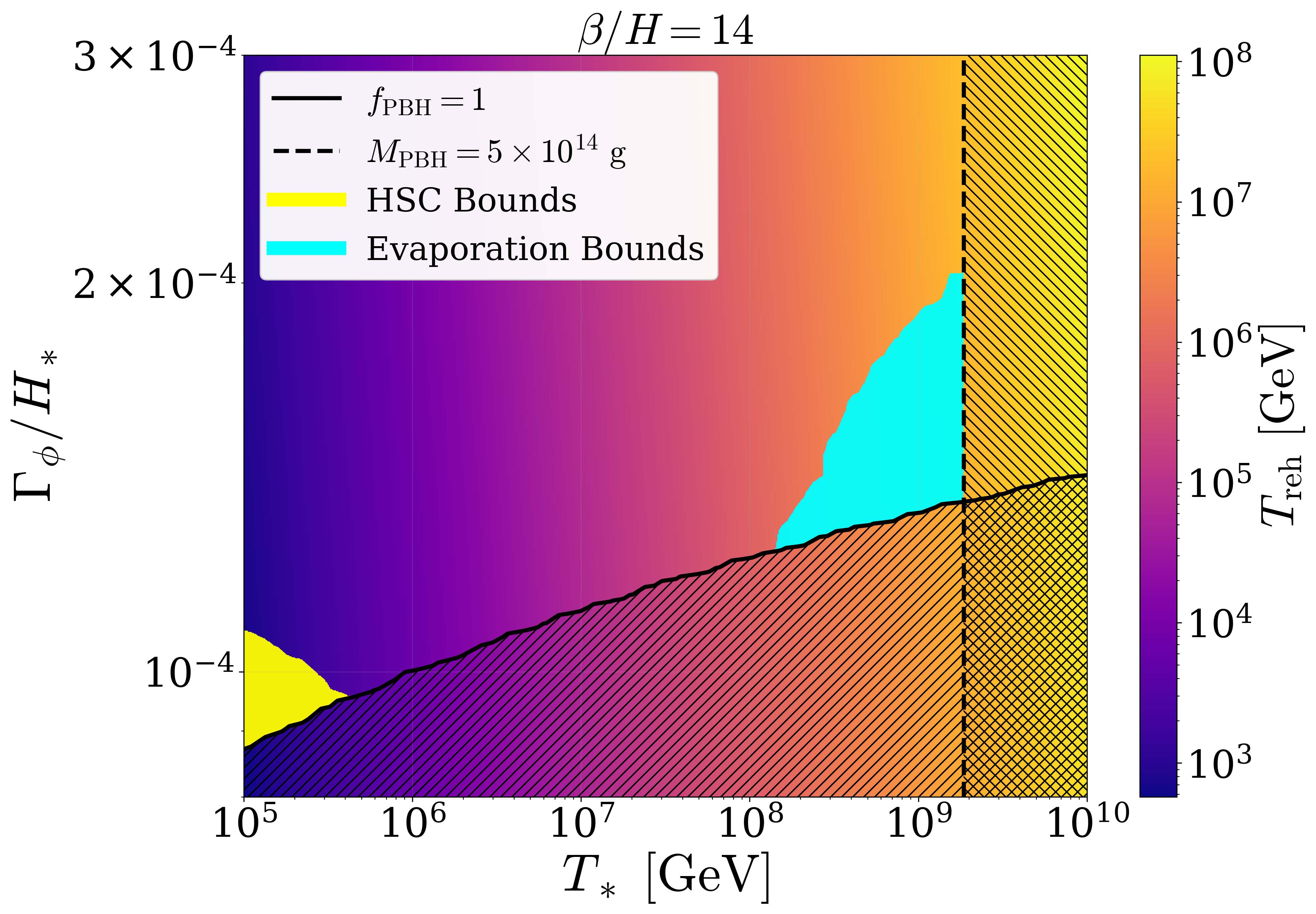}
\caption{Constraints on $T_* - \Gamma_{\phi}/H_*$ plane from (non-)observation of PBHs in the form of PBH evaporation~\cite{Carr:2009jm, Zhang:2007zzh, Adams:1998nr, EGRET:1997qcq, Wright:1995bi, MacGibbon:1991vc, Boudaud:2018hqb} and microlensing from HSC Subaru~\cite{Niikura:2017zjd} for $\beta/H = 8,~14$. The black solid line corresponds to $f_{\rm PBH} = 1$ and the vertical dotted line corresponds to the lightest non-evaporated PBH mass, i.e., $M_{\rm PBH} = 5\times 10^{14}\mathrm{~g}$. The color coordinate denotes the reheating temperature resulting from the combination of $T_* - \Gamma_{\phi}/H_*$.}
\label{fig:pbhabun}
\end{figure}

In the previous subsection, we discussed the initial collapse fraction of the PBHs. However, as explained in the previous section, the present observable is the PBH abundance. The present PBH abundance depends on $\beta/H$, $\Gamma_{\phi}/H_*$ and $T_*$. Therefore, in this section we show existing (non-)observational bounds on PBH abundance in the $(T_*) - (\Gamma_\phi/H_*)$ plane for fixed $\beta/H$ values. Since most of these bounds are based on the assumption that PBHs are monochromatic and our relevant PBH mass range only spans roughly by a factor of two, we consider $f_{\rm PBH} = \int d(M/M_{\rm min})f(M/M_{\rm min})$ and $M_{\rm PBH}$ to be the mass where $f(M/M_{\rm min})$ peaks. Furthermore, we consider the range of $T_*$ and $\Gamma_{\phi}/H_*$ such that $T_{\rm reh} \gtrsim 10^3\mathrm{~GeV}$ so that the reheating happens before the electro-weak phase transition. Within this domain only the bounds from PBH evaporation and some parts of PBH microlensing are applicable, which we show for two representative values of $\beta/H = 8,~14$ in Fig.~\ref{fig:pbhabun}.
The PBH mass scale depends almost entirely on $T_*$; therefore, the dotted line in the above figure, which corresponds to the lightest non-evaporated PBH mass, corresponds to $T_* \sim 2\times 10^{9}\mathrm{~GeV}$. On the other hand, the PBH abundance has a strong dependence on both $T_*$ and $\Gamma_{\phi}/H_*$ where the former arises from the redshift post reheating and the latter results from the behaviour of initial collapse fraction on the reheating time. Therefore, the solid black curve, which denotes the lower bound of $\Gamma_{\phi}/H_*$ arising from the PBH over-abundance is slanted heavily. The color coordinate denotes the reheating temperature and in both cases our parameter domain is such that the reheating temperature is larger than the electro-weak phase transition temperature. Although in principle $T_{\rm reh}$ depends on both $T_*$ and $\Gamma_{\phi}/H_*$, in these plots the former spans many orders of magnitude whereas the range of the latter is very narrow, highlighting only $T_*$ dependence.  

Since PBH masses are negatively correlated with $T_*$, the evaporation bounds (cyan region), which are sensitive to extremely light PBHs are situated at high values of $T_*$ ($> 10^8\mathrm{~GeV}$). On the contrary, the HSC microlensing bounds (yellow region) sensitive to (sub-) lunar mass PBHs are at the lower values of $T_*$ ($< 10^6\mathrm{~GeV}$). The region where $T_* \in (10^6,10^8)\mathrm{~GeV}$ is only constrained by PBH-over abundance bounds and hence FOPTs with properties similar to this region of the parameter space can source PBHs which can play the role of the dark matter in the universe entirely.

The two representative values of $\beta/H$ also shows that although the $T_*$-related aspects of the plots are independent of $\beta/H$, there exists a trade-off between $\beta/H$ and $\Gamma_{\phi}/H_*$: higher values of $\beta/H$ allow lower values of $\Gamma_{\phi}/H_*$. This is a reflection of the behaviour of initial collapse fraction with varying $\beta/H$ and $\Gamma_{\phi}/H_*$.

In the previous subsection we discussed the exponential sensitivity of the initial collapse fraction on $\Gamma_{\phi}/H_*$; this sensitivity also translates to $f_{\rm PBH}$ as can be seen for the above figure. More importantly, to the best of our knowledge this study provides the first bounds on the reheating efficiency associated with FOPT-driving scalars from PBH abundance. Specifically we can see that in the allowed region, for $\beta/H = 8,~14$ one requires $\Gamma_{\phi}/H_* \in (4.2\times 10^{-4}, 4.8\times 10^{-4})$, and $(9\times 10^{-5}, 1.1\times 10^{-4})$ respectively to achieve $f_{\rm PBH} = 1$.

\section{Summary and Conclusion}
\label{sec:conc}
Recently, slow FOPTs were revived as a PBH-generating mechanism, provided that there is an EMD phase between percolation and reheating. The duration of this phase is governed by the ratio of the decay rate of the FOPT-driving scalar and the Hubble parameter at percolation, i.e., $\Gamma_{\phi}/H_*$. In this article, we study this mechanism in detail to obtain the role of $\beta/H$, $\Gamma_{\phi}/H_*$, and $T_*$ on the present PBH abundance. Furthermore, we investigate the overdensity threshold for collapse in EMD and the initial spin distribution of the PBH population.

Using the non-Gaussian PDF of density perturbations generated during a slow FOPT, we constructed an effective power spectrum for the same. We obtained the spectral moments of the effective power spectrum for various values of $\beta/H$ and $k/k_{\rm max}$. We employed a modern peak-theory approach to study the collapse of overdensities during EMD. We included the effects of the full deformation tensor, including the off-diagonal elements from the tidal torque and the Hessian eigenvalues. Since current simulations do not measure the distributions of these quantities arising from FOPT, we employed an effective approach to construct the joint distribution of these peak variables: we used the non-Gaussian distribution of the overdensity, assuming that the other peak variables follow a joint Gaussian distribution. At this stage, we ran a large-scale Monte Carlo simulation for different values of the peak variables at each point in a dense grid of the zeroth moment of the effective density power spectrum and the overdensity. For each realization at each grid point, we checked whether the resulting collapsed objects satisfied the hoop conjecture, which eventually allowed us to obtain the collapse probability, spin distribution, collapse time, etc.

We find that, unlike in RD, in FOPT followed by EMD, the collapse threshold emerges dynamically through the competition between the collapse and reheating times. Therefore, the initial collapse fraction of the PBHs is extremely sensitive to the factor $\Gamma_{\phi}/H_*$; a change of a factor of two in this quantity can lead to a change of tens of orders of magnitude in the collapse fraction. The collapse fraction is also strongly anti-correlated with $\beta/H$. Finally, we found that even after including tidal-torque effects following a modern peak-theory formalism, the average spin parameter for all cases is $\mathcal{O}(10^{-3})$. This is because the hoop conjecture criterion filters out high angular-momentum regions. We also translate the PBH (non-) observational bounds arising from evaporation and microlensing on the $T_* - \Gamma_{\phi}/H_*$ plane. We find that for $\beta/H \sim \mathcal{O}(10)$ and $\Gamma_{\phi}/H_* \sim \mathcal{O}(10^{-4})$, resulting PBHs can play the role of the dark matter of the universe in its entirety.

We emphasize that we considered two assumptions: (i) in the absence of a density power spectrum, we constructed an effective one to obtain the local properties of the density perturbation in $k-$space, and (ii) due to the limitation of the present simulations to generate the distribution of peak variables other than $\nu$, we consider a Gaussian peak theoretic joint distribution for these quantities. With targeted future simulations, one can relax both of these assumptions; however, this work provides a first peak-theoretic estimate of the properties of PBHs arising from slow FOPTs with delayed reheating. Furthermore, if the EMD phase is sufficiently prolonged, PBHs formed from highly overdense regions (which collapse at earlier times) may undergo significant accretion. However, we do not consider accretion in the present work, since we find that such long-lived EMD phases generally lead to an overproduction of PBHs and are therefore observationally disfavored.

In conclusion, this study connects FOPT and reheating parameters to the observable properties of the resulting PBH population. Future detection (non-observation) of PBHs can shed light (place bounds) on the underlying microphysics of the early universe.

%
 

\bibliographystyle{JHEP}
\bibliography{Ref.bib}

\end{document}